# Formation of nickel clusters wrapped in carbon cages: towards new endohedral metallofullerene synthesis

*Alexander S. Sinitsa[1], Thomas W. Chamberlain[2], Thilo Zoberbier[3], Irina V. Lebedeva[4,\*], Andrey M. Popov[5,\*], Andrey A. Knizhnik[1,6], Robert L. McSweeney[7], Johannes Biskupek[3], Ute Kaiser[3], Andrei N. Khlobystov[7]*

[1] National Research Centre "Kurchatov Institute", Kurchatov Square 1, Moscow 123182, Russia

[2] Institute of Process Research and Development, School of Chemistry, University of Leeds, Leeds, LS2 9JT, UK

[3] Group of Electron Microscopy of Materials Science, Central Facility for Electron Microscopy, Ulm University, Albert Einstein Allee 11, Ulm 89081, Germany

[4] Nano-Bio Spectroscopy Group and ETSF, Universidad del País Vasco, CFM CSIC-UPV/EHU, San Sebastian 20018, Spain

[5] Institute for Spectroscopy of Russian Academy of Sciences, Fizicheskaya Street 5, Troitsk, Moscow 108840, Russia

[6] Kintech Lab Ltd., 3rd Khoroshevskaya Street 12, Moscow 123298, Russia

[7] School of Chemistry, University of Nottingham, University Park, Nottingham NG7 2RD, UK

ABSTRACT In spite of the high potential of endohedral metallofullerenes (EMFs) for application in biology, medicine and molecular electronics and recent efforts in EMF synthesis, the variety of EMFs accessible by conventional synthetic methods remains limited and does not include, for example, EMFs of late transition metals. We propose a method in which EMF formation is initiated by electron irradiation in aberration-corrected high-resolution transmission electron spectroscopy (AC-HRTEM) of a metal cluster surrounded by amorphous carbon inside a carbon nanotube serving as a nano-reactor and apply this method for synthesis of nickel EMFs. The use of AC-HRTEM makes it possible not only to synthesize new, previously unattainable nanoobjects but also to study in situ the mechanism of structural transformations. Molecular dynamics simulations using the state-of-the-art approach for modeling the effect of electron irradiation are performed to rationalise the experimental observations and to link the observed processes with conditions of bulk EMF synthesis.

Endohedral metallofullerens (EMFs) and their derivatives have shown considerable promise for biological and medical applications and for molecular electronics[1,2,3]. However, the variety of EMFs available by current synthetic methods is very limited. Traditional methods for the synthesis of EMFs in the arc discharge reactor or via laser evaporation are restricted principly to alkali earth elements, lanthanides and early transition metals (groups 2 – 4 of the Periodic Table), see Refs. 1 – 3 for a review. Along with conventional EMFs, clusterfullerenes with endohedral metal nitride[4], carbide[5], and other metal compounds have been produced via modified arc discharge methods enabling the addition of a select group of small molecules. Endohedral non-metallic fullerenes with endohedral atoms of noble gases and nitrogen have also been produced by treatment of empty fullerenes after their synthesis, for example by high-pressure incorporation[6]. Endohedral non-metallic fullerenes with very small endohedral molecules have also been obtained via long sequences of organic reactions[7,8]. There is also a limit to the number of atoms which can be contained inside fullerene cages, with 7 atoms[9] for $Sc_4O_3@C_{80}$ the maximum to date. Thus elaboration of methods of EMF production which enables extention of the EMF family to late transition metals and/or to increase the size of the complexes inside the fullerene cage is extremely timely.

A new method to produce and image fullerenes as a result of graphene flake transformation under electron irradiation in AC-HRTEM has been elaborated recently[10]. Such a method opens up possibilities to produce new types of EMFs by electron irradiation treatment of nanostructures starting from a metal cluster attached to sufficient carbon material, i.e. about 100 carbon atoms. In contrast to the formation of EMFs at high temperature in the arc discharge reactor or by laser evaporation, where a weakly bonded metal atom or cluster can be easily removed from the system through sintering or vapourisation, transformations under electron irradiation take place at room temperature where electron impact is powerful enough to remove light carbon atoms but heavy metal atoms are conserved in the system. Here we obtain EMFs consisting of nickel clusters of several tens of atoms inside the fullerene cage by electron irradiation treatment of nickel clusters surrounded by amorphous carbon or a graphene flake during AC-HRTEM imaging. The proposed method can be extended to any metal cluster of a given composition and number of atoms including clusters composed from several chemical elements.

Metals in groups 9 – 11 of the Periodic Table are very important in the context of catalysis of carbon nanostructure growth, such as carbon nanotubes and graphene. Therefore, obtaining a detailed understanding of nickel cluster EMF formation can provide an insight into the nucleation of $sp^2$ hybridized carbon on the catalyst particle, a starting point which is known to be crucial in determining the structure and property of carbon nanotubes for example. To study the formation of Ni EMFs in detail we utilised confinement inside carbon nanotubes, a new important class of nano-reactors, that represent an inert and controlled reaction environment whilst the atomically thin sidewalls enable in situ AC-HRTEM imaging. TEM is the only tool that enables direct investigation of the structure and properties of metal clusters encapsulated in the nanotube. When the energy and dose rate of e-beam are controlled, it is possible to trigger chemical reactions between metal clusters and carbon. HRTEM imaging reveals important dynamics of metal-carbon bonding and a wide range of chemical processes catalyzed by metals under the influence of the kinetic energy of the e-beam have been observed with atomic resolution.[11-16] The crucial remaining challenge is to link these observed processes with conditions of preparative reactions for carbon nanostructure growth (e.g. fullerene formation in electric arc discharge, nanotube growth in CVD processes etc), which requires modelling and systematic comparison of processes under controlled e-beam irradiation and heating conditions.

Let us first discuss the experimental details of synthesis of nickel-carbon heterostructures. Clusters of nickel metal were encapsulated in single-walled carbon nanotubes (SWNTs) in the form of nickel hexafluoroacetylacetonate, $Ni(C_5HF_6O_2)_2$, which can be easily broken down into pure metal and ligand. While the nickel atoms aggregate into clusters of 50-100 atoms forming intimate contact with the nanotube inner (concave) surface, labelled NiNPs@SWNT, the ligand is broken into small fragments and leaves the nanotube. SWNTs (arc-discharge, NanoCarbLab) were annealed at 540 °C for 20 minutes to open their termini and remove the majority of residual amorphous carbon from the internal cavities, a 20% weight loss was observed. Nickel hexafluoroacetylacetonate (10 mg) (used as supplied, Sigma Aldrich) was mixed with the SWNT (5 mg), sealed under vacuum ($10^{-5}$ mbar) in a quartz ampoule and heated at 140 °C, a temperature slightly above the vaporisation point of the metal complex, for 3 days to ensure complete penetration of the SWNT. The sample was then allowed to cool to room temperature, washed repetitively with tetrahydrofuran to remove any metal complex from the exterior of the SWNT and then filtered through a PTFE membrane (pore diameter = 0.2 µm). The nanotubes sample filled with metal complex was then sealed in a quartz ampoule under an argon atmosphere and heated at 600 °C, a temperature significantly above the decomposition point of the metal species (~150-200 °C), for 2 hours to decompose the metal complex into the desired pure metal nanoparticles.

Alternatively, the decomposition process can be achieved directly during TEM using the e-beam as the energy source. Metal particles formed by thermal and e-beam decomposition of the metal complex are virtually indistinguishable.

AC-HRTEM was used to verify the presence of nickel clusters inside the carbon nanotubes (Fig. 1a) and the identity of the metallic clusters formed inside the nanotubes was confirmed by energy dispersive X-ray (EDX) spectroscopy (Fig. 1b). EDX spectra were recorded for small bundles of SWNTs (3-10 nanotubes) filled with nickel metal on a JEOL 2100F TEM equipped with an Oxford Instruments X-rays detector at 100 kV. Evolution of the metal clusters and their interactions with elemental carbon were analyzed by AC-HRTEM imaging (see Fig. 1c-e and Figs. S1-3 in Supporting Information) carried out using an image-side $C_S$-corrected FEI Titan 80-300 transmission electron microscope operated at 80 kV acceleration voltage with a modified filament extraction voltage for information limit enhancement and increased image contrast[17]. Images were recorded on a slow-scan CCD-camera type Gatan Ultrascan XP 1000 using binning 2 (1024 by 1024 pixel image size) with exposure times between 0.25-1.0 s. For all in situ irradiation experiments the microscope provided a highly controlled source of local and directed electron irradiation on a selected area of the sample. Experimentally applied electron-fluxes ranged from $1 \cdot 10^6$ to $1.5 \cdot 10^7$ e$^-$/(nm$^2 \cdot$s), and the total applied dose was kept the same, reaching up to $5 \cdot 10^9$ e$^-$/nm$^2$ at the end of each experiment. TEM specimens were heated in air at 150 °C for about 5 min shortly before insertion into the TEM column. All imaging experiments were carried out at room temperature. Subsequent TEM image simulations (see Fig. 1f and Fig. S4 in Supporting Information) were performed for structure models based on molecular dynamics (MD) calculations (see Fig. 1g and details about MD later in the text) using the multi-slice program QSTEM developed by C.T. Koch[18]. Image simulation parameters were: 80 keV electron energy, spherical aberration parameter Cs = 10μm, focus = -8 nm (corresponds to Scherzer-focus conditions with black atom contrast) and focus spread = 4 nm. The dampening effect of the camera (MTF, modulation transfer function) was included. The effect of limited electron dose was emulated by applying noise to the calculated (infinitive dose) images using a custom-made Monte-Carlo program exploiting the Poisson statistics of electrons.

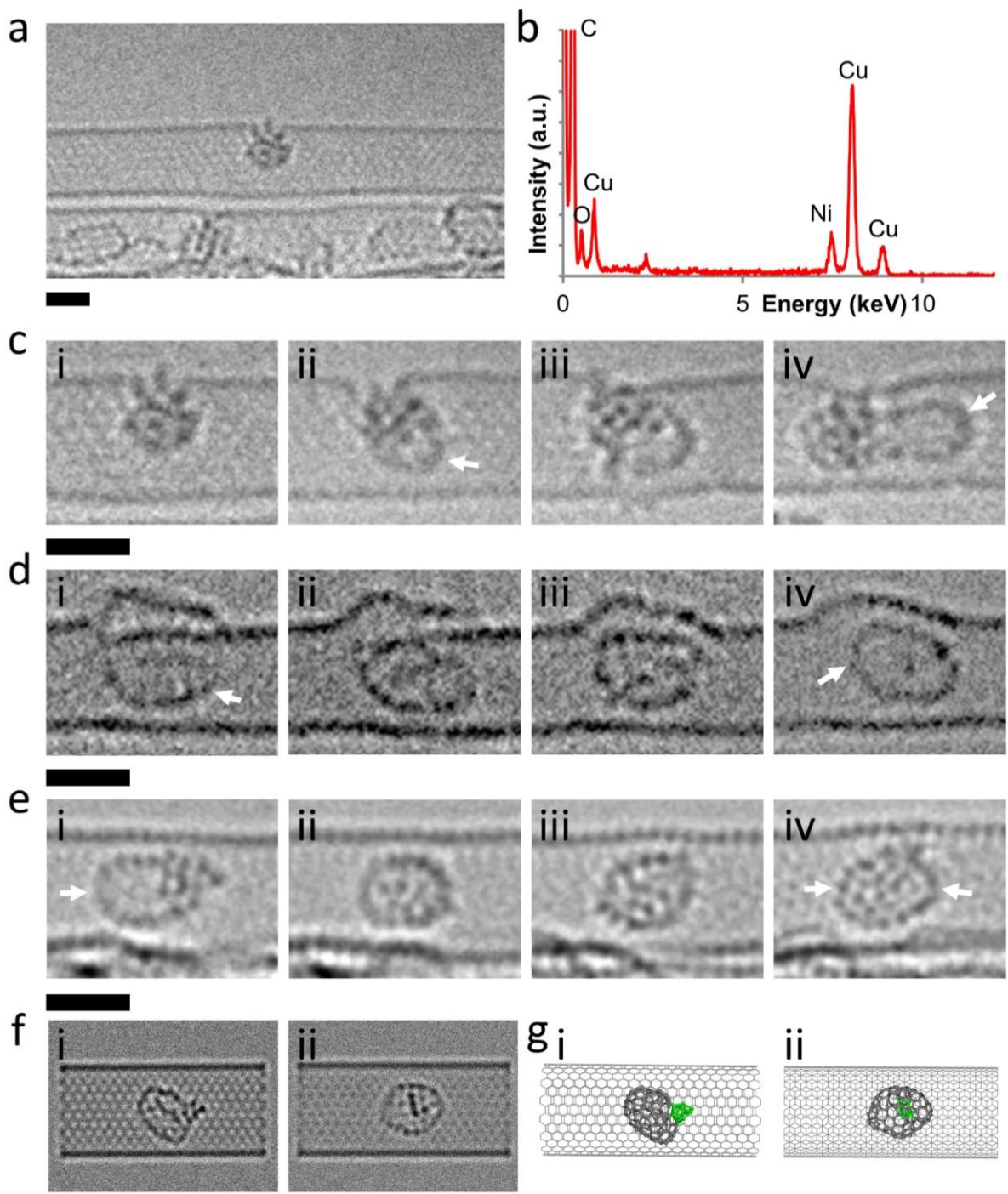

**Figure 1.** Structural evolution of nickel clusters with carbon over time under 80 keV electron irradiation: (a) AC-HRTEM image of NiNPs@SWNT showing a cluster of 50-70 nickel atoms encapsulated inside a SWNT; (b) Energy dispersive X-ray spectroscopy of the NiNPs@SWNT sample confirms the presence of nickel inside the nanotubes (Cu peaks are due to the TEM grid). Time series of experimental AC-HRTEM images showing the key stages of nickel cluster carbon

transformation, (white arrows depict carbon cages in all cases): (c) (i-iv), shows evolution of a patched fullerene in which the nickel cluster interacts with carbon atoms of the nanotube sidewall and extrudes carbon in the form of a fullerene cage whilst simultaneously becoming incorporated within the cage structure; (d) (i-iv) and (e) (i-iv), depict examples of endohedral metallofullerene formation in which carbon cages fully encase nickel clusters of various sizes, passivating the nickel and preventing it from interacting with the nanotube sidewalls. Strong correlation is observed between the contrast in the simulated TEM images (f) (i and ii) and the corresponding experimental AC-HRTEM images (e) (i and ii) respectively. Simulated images were generated by applying a limited dose of 80 keV electrons to the structural models depicted in (g) (i-ii). The scale bar is 1 nm in all cases.

Time series of images recorded for individual nickel clusters consistently show interactions between the clusters and carbon, both from the nanotube sidewall and amorphous carbon inherently present inside carbon nanotubes (Fig. 1c-e). The energy of the electron beam (80 keV) in the AC-HRTEM experiment was carefully selected to enable the evolution of the carbon/nickel cluster system to be observed over extended time periods whilst simultaneously providing sufficient energy to initiate transformations without irreversibly damaging the resultant structures. The metal cluster remains highly flexible, changing its shape continuously during the AC-HRTEM observation, (Fig. 1c-e) which implies low-directional and flexible metallic bonding between nickel atoms, rather than a covalent or ionic bonds that would lead to a static, highly ordered cluster, such as iron carbide under the same experimental conditions[19]. Furthermore, the distance of (2.2 ± 0.1) Å that can be clearly measured between nearest neighbour nickel atoms in some configurations of the cluster (Fig. 1c, for example) matches the distance between projections of atoms onto the (111) plane for metallic nickel (see modeling part of this study). These two facts confirm that nickel remains in the metallic form under our experimental conditions.

Carbon being a low atomic number element (Z = 6 for carbon as compared to Z = 28 for nickel) has a lower single-atom contrast in AC-HRTEM images than nickel, as can be clearly seen by comparing the contrast of the carbon nanotube with the nickel cluster (Fig. 1a,c). Analysis of the time-series images reveals a growth of new low-contrast structures on the nickel surface (Fig 1c-e), which are not part of the initial nickel cluster, and result from interactions between nickel and carbon nanotube under the e-beam. The carbon nature of these structures is supported by the comparison of the experimental AC-HRTEM and simulated TEM images (Fig. 1e-g and Fig. S4 in Supporting Information). The carbon features either extend from (Fig. 1c) or completely encase (Figs. 1d and e) the nickel clusters, resembling elongated hemispherical structures similar to nanotubes or fullerene-like carbon cages, respectively. Our time-resolved AC-HRTEM images clearly indicate that the nickel clusters act as a stable surface for the carbon cages to evolve, and provide a source of dissolved carbon atoms for the cage to grow (Fig. 1c). The formation of cage-like structures from amorphous carbon is driven by the energetic requirement of the carbon system to reduce the number of high energy, low coordinate atoms in similar fashion to the previously reported example in which a graphene sheet rolls up to form $C_{60}$ under e-beam irradiation conditions[10]. The nickel clusters appear to be intimately incorporated in the structure of the carbon cage, acting as a patch in the structure during all initial time series before being slowly encased by the carbon cage with the cluster acting as a template for fullerene formation, finally resulting in the formation of structure where the metal is encased in the carbon shell.

Our experimental measurements clearly indicate (Fig. 1d and e) that the characteristic diameter of the carbon cages formed is about 1 nm. Using the empirical formula[20] $d \approx 0.71 \times (N/60)^{1/2}$ that links the average fullerene diameter, $d$, and the number of carbon atoms in the fullerene cage, $N$, we estimate that carbon cages formed comprise 100 – 200 atoms. In literature carbon cages up to $C_{2160}$ are considered as fullerenes[21]. Fullerenes with sizes up to $C_{330}$ have been synthesized experimentally[22]. Large endohedral fullerenes such as $La_2C_{138}$ can also exist [23]. Therefore the carbon cages with metal clusters inside observed in our study can be referred to as EMFs.

The process of EMF formation is dependent on both being sufficient carbon to encase the entire cluster and also being enough space inside the nanotube to enable the cage to fit around the nickel cluster. Upon formation of the EMF the resultant encased nickel cluster is passivated and no further interactions are observed with carbon atoms within the nanotube. For larger clusters with insufficient room in the nanotube for the cage to fit around the metal cluster, the carbon shell is observed to grow from the nickel cluster, with the cluster acting as a support/surface and being incorporated in the carbon framework, forming so-called patched heterofullerenes (Fig. 1c), which is structurally related to a short open-ended nanotube.

The mechanisms of EMF formation are complex and not fully understood as the synthesis traditionally takes place under harsh conditions (e.g. high temperature, electric discharge etc.) making it impossible to follow the formation mechanisms using current analytical methods. Our AC-HRTEM images capture the key stages of the nucleation, growth and rearrangement of the carbon cages interacting with nickel, which can shed light on the actual mechanisms of endohedral fullerenes formation and potentially open avenues for new types of EMFs. However, as AC-HRTEM technology is still unable to reveal the full atomistic mechansism of this process, it is important to employ theoretical modeling. Thus MD simulations have been performed to study the atomistic mechanisms of formation of EMF with 1-2 metal atoms inside carbon cage starting from metal-carbon vapor[24,25] and reactions of $C_2$ insersion into EMF cage.[26] Since the development of the CompuTEM algorithm[27,28], it is possible to realistically model chemical reactions induced by the e-beam. The high-temperature MD step of this algorithm that is introduced after electron collisions that have resulted in structural transformations provides efficient description of structure relaxation that under real experimental conditions takes place in seconds. In this study we use a version of this approach in which the momentum transferred from incident electrons to nuclei is treated explicitly and no *a priori* information on possible reactions and their cross-sections is needed[28]. Previously MD simulations using CompuTEM algorithm[27,28] have been applied to study graphene-fullerene transformation[27,28], formation of metal heterofullerenes[29] and the cutting of carbon nanotubes by nickel clusters[15] under electron irradiation.

To explore and deconvolute the separate roles of the electron irradiation and carbon nanotube confinement, three cases of treatment have been considered: electron irradiation inside a carbon nanotube, heat treatment inside a carbon nanotube, and electron irradiation in vacuum. Since thermodynamical fluctuations have considerable influence on nonequilibrium processes in small systems, 50 simulation runs were performed for each of these cases.

The initial structure of amorphous carbon consisting of 160 atoms and surrounding the nickel cluster of 13 atoms was prepared by sequential addition of carbon dimers at random points located at distances between 1.4 Å to 5 Å from the cluster center followed by annealing at high temperature ($T_{ann} = 2500$ K). To mimic the structures observed in AC-HRTEM (Fig. 1d), where a nanotube serves as a reactive container, the cluster surrounded by amorphous carbon was inserted inside the (14,14) carbon nanotube with a hemispherical protrusion of 14.6 Å in diameter. No covalent bonds were formed between the amorphous carbon structure or nickel cluster and the carbon nanotube wall as the distance between them during MD runs normally exceeded 3 Å.

The nickel-carbon potential developed in our previous papers[29,38] was used to describe interactions between the amorphous carbon and nickel cluster within these structures. This potential was elaborated on the basis of the first-generation bond-order Brenner potential and reproduces both the experimental and first-principle data on the physical properties of metallic nickel as well as the relative energies of carbon species on nickel surfaces and in the bulk nickel metal.[38] The potential was also previously applied for simulation of carbon nanotube cutting catalyzed by nickel under electron irradiation showing the qualitative agreement of structure evolution with the experimental observations and correctly predicting the key stages of the cutting.[15] The carbon nanotube structure was geometrically optimized and then fixed. The Lennard-Jones 12-6 potential $U(r) = 4\varepsilon \left[ (\sigma/r)^{12} - (\sigma/r)^{6} \right]$ was used for interactions between the carbon nanotube wall and all atoms inside it. The parameters for interactions between the amorphous carbon and carbon nanotube ($\varepsilon_{CC} = 3.40$ meV, $\sigma_{CC} = 3.73$ Å) were taken from the AMBER force-field for aromatic carbon atoms[30]. The parameters for interactions between the nickel cluster and the carbon nanotube ($\varepsilon_{CNi} = 5.28$ meV, $\sigma_{CNi} = 1.78$ Å) were chosen to reproduce the binding energy and equilibrium distance for a graphene layer on the nickel (111) surface obtained on the basis of the adiabatic-connection fluctuation-dissipation theorem in the random phase approximation (RPA)[31,32]. This approach offers a natural way to describe non-local van der Waals interactions and is much more accurate compared to density functional theory methods even with special van der Waals functionals. In particular, the RPA method gives a value for the distance between the graphene layer and the nickel surface that is very close to the experimental value of $r_0 = 2.11$ Å (see Ref. 33 for review). Average values of the binding energy and the equilibrium distance from two RPA calculations[31,32] were used as the fitting parameters of the potential ($E_{gr-Ni} = -68.5$ meV and $r_0 = 2.18$ Å, respectively).

MD simulations were carried out using the in-house MD-kMC[34] (Molecular Dynamics – kinetic Monte Carlo) code. The integration time step was 0.6 fs. In simulations without electron irradiation, the temperature was maintained at $T = 2500$ K by the Berendsen thermostat[35] with a relaxation time of 0.3 ps. In simulations with electron irradiation, the electron flux was $j = 4 \cdot 10^6$ e$^-$/(nm$^2 \cdot$s) and the kinetic energy of electrons was $E = 80$ keV. To exclude thermally induced structural transformations during the high temperature MD step, the temperature and duration of this step were chosen to be $T_{el} = 1800$ K and 100 ps, respectively[29]. The relaxation times of the thermostat were 0.1 ps, 3 ps and 0.3 ps for steps 1, 5 and 6 of the CompuTEM algorithm[27,28], respectively. The criteria used to distinguish the nanoobjects formed (EMF, patched heterofullerene, fullerene with attached nickel cluster) are described in Note S1 in Supporting Information.

The simulations for all three cases considered (electron irradiation inside the carbon nanotube and in vacuum and heat treatment inside the carbon nanotube) revealed qualitative similarities in terms of the specific processes of EMF and heterofullerene formation. Transformation of amorphous carbon into a fullerene cage takes place with the resulting formation of a heterofullerene with a nickel patch or an EMF with a nickel core depending on the position of the nickel cluster at the moment when the fullerene cage builds up. Examples of the MD simulations carried out to explore the electron irradiation intitated tranformations inside the carbon nanotube resulting in EMF and into a patched heterofullerene formation are shown in Figs. 2a and 2b, respectively. Fig. 2a shows fast formation of the closed fullerene cage around the nickel cluster and the stability of this cage during the simulation time. The formation time was in the order of 1000 s and the accumulated dose of $4 \cdot 10^9$ e$^-$/nm$^2$ in the MD simulations correlates very well with the characteristic time scales and doses observed in the experiments. The average interatomic distance between nickel atoms in the cluster is 2.4 Å, which corresponds to the distance between nearest neighbour projections of atoms onto the (111) plane of 2.1 Å, in agreement with the experimental observations. The presence of structural defects like heptagons in the fullerene cage is typical for MD simulations of fullerene formation due to the relatively small simulation time accessible for MD, see, for example, simulations of formation of fullerenes from carbon vapor[36] and EMFs with one-two metal atoms inside the carbon cage from metal-carbon vapor[24,25]. Examples of structure evolution simulations which form the same nanoobjects under electron irradiation in vacuum are shown in Fig. S5 in Supporting Information. The more perfect structure of EMFs formed in vacuum compared to the ones inside the nanotube is related to an increase in the accessible simulation time due to the decrease in size of the simulated system. The observation of carbon chains attached to the fullerene cage are also typical in simulations of fullerene formation[36].

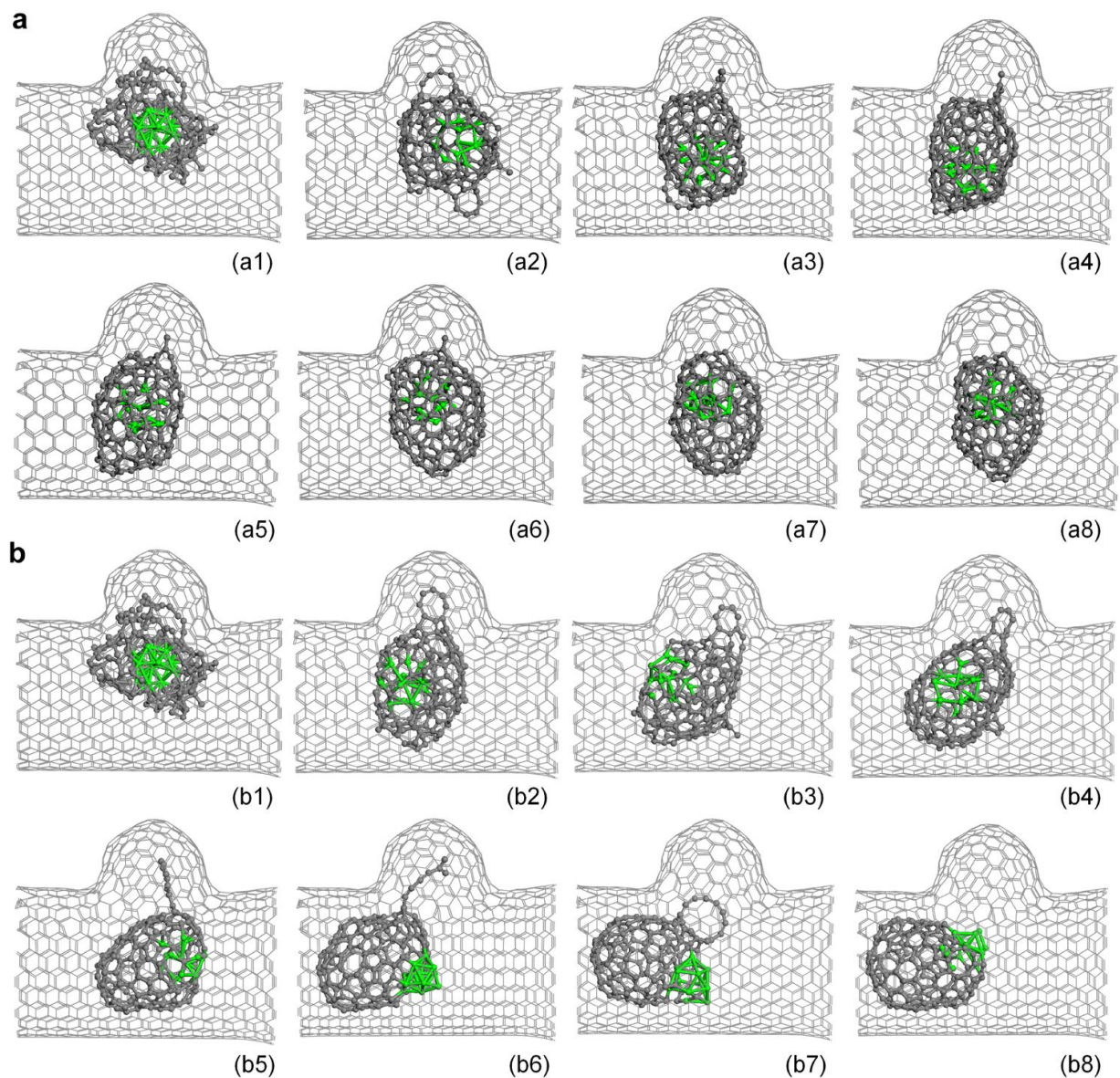

**Figure 2.** Structure evolution of nickel clusters with carbon in molecular dynamics simulations under electron irradiation: (a)-(b), Simulated structure evolution of the amorphous carbon with a nickel cluster attached inside the carbon nanotube under electron irradiation in AC-HRTEM: (a) transformation into an endohedral metallofullerene, structures observed at: (a1) 0 s, (a2) 50 s, (a3) 149 s, (a4) 300 s, (a5) 450 s, (a6) 650 s, (a7) 750 s and (a8) 900 s; b, transformation into a patched heterofullerene, structures observed at: (b1) 0 s, (b2) 50 s, (b3) 152 s, (b4) 200 s, (b5) 450 s, (b6) 602 s, (b7) 750 s and (b8) 900 s.

In all three considered cases of amorphous carbon treatment EMF formation is observed for only part of the simulation runs and thus can be classified as a casual event in accordance with our experimental observations. Note that only patched heterofullerenes were observed in simulations exploring the transformation of a graphene flake with a nickel cluster attached under electron irradiation in vacuum.[29] This implies that for EMF formation from a carbon nanostructure with a metal cluster attached, amorphous carbon is the preferable starting structure compared to graphene flakes. EMFs are found to be significantly more stable under electron irradiation and heat treatment than patched heterofullerenes, which transform into closed fullerene cages with the nickel clusters attached through a relatively small number of bonds to the outside of the cage and typically results in desorption of the nickel clusters over time. This qualitative conclusion on the relative stability of EMFs and patched heterofullerenes is in agreement with the greater binding energy for the most part of EMFs in comparison with patched heterofullerenes (Fig. 3a).

Although common qualitative features are observed for EMF and patched heterofullerene formation in the three considered cases of treatment, time dependences of fractions of different nanoobjects shown in Fig. 4 reveal considerable differences in EMF and patched heterofullerene maximal yields (Fig. 3b). The yields of EMFs and heterofullerenes under electron irradiation compared to heat treatment are signifiacntly greater (observed increases of 600 % and about 150 % for EMFs and heterofullerenes, respectively). This differs from general expectations that transformation processes should occur similarly for any source of energy for bond rearrangements, i.e. heat treatment or electron impacts. We believe that the latter holds for pure carbon systems. For example, in the experimental studies of graphene nanoribbon formation inside carbon nanotubes[37] and simulations of the graphene-fullerene transformation[28,38] no difference was found between the products of heat treatment and electron irradiation. However in metal-carbon systems the energy received from electron impacts is transmitted mainly to light carbon atoms. Therefore electron irradiation promotes mainly bond rearrangement within the carbon network. This is in stark contrast to heat treatment, which favours penetration of the nickel cluster through the fullerene cage and desorption of the cluster from the cage. Interestingly approximately 4 carbon atoms per simulation are removed from the system in the simulations under electron irradiation inside the carbon nanotube. Since carbon atoms bonded to nickel atoms have a considerably greater probability of being removed from the metal-carbon system by electron impacts than three-coordinated atoms of the carbon network,[15] knock-out of such atoms gives some preference to formation of EMFs in comparison to heterofullerenes. This effect can also contribute to the greater EMF yield under electron irradiation.

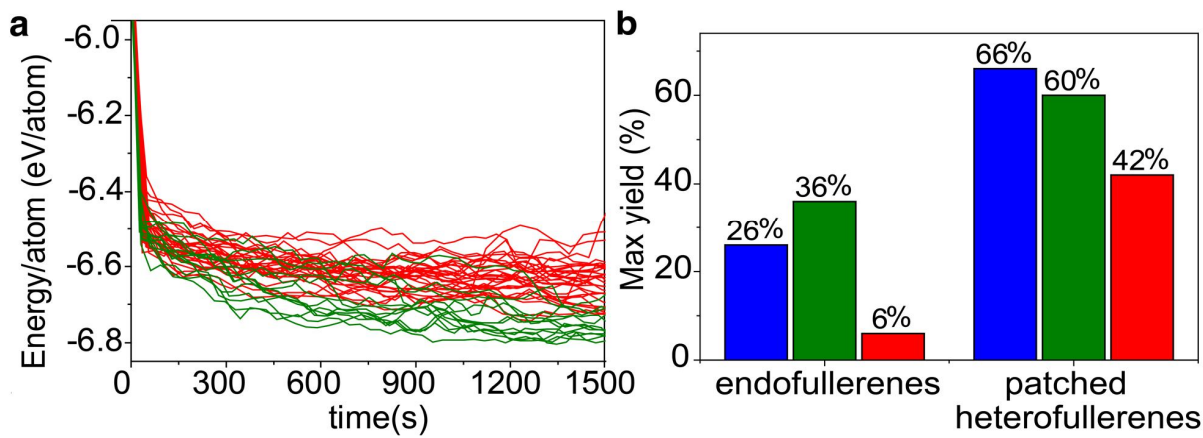

**Figure 3.** Time dependences of the potential energy and maximum yield of endo- and heterofullerenes observed in molecular dynamics simulations: (a) Calculated time dependences of the potential energy per atom of the nickel-carbon structure under electron irradiation in AC-HRTEM in vacuum for 50 molecular dynamics runs. The results of the simulations where endohedral metallofullerenes and patched heterofullerenes were formed are shown by green and red lines, respectively; (b) Calculated maximum yields of endohedral metallofullerenes and patched heterofullerenes under electron irradiation in AC-HRTEM in vacuum (blue bars), under electron irradiation inside the carbon nanotube (green bars) and under heat treatment inside the carbon nanotube (red bars).

As the EMF yield under electron irradiation inside the carbon nanotube is about 30% greater than in vacuum it is clearly important to consider also the possible role of the nanotube in such transformations. The transformation of amorphous carbon into a fullerene cage is a directed process which occurs with a considerable decrease of the total energy shown in Fig. 3a and without reverse transitions (see Figs. 4d, e and f and experimental observations (Fig. 1)). The influence of the nanotube on this transformation is not essential. In contrast penetration of the nickel cluster through the fullerene cage during the cage formation and thereafter is possible in both directions, into and out of the cage, through a hole in the cage that is not completely closed yet or through a hole formed in the closed fullerene cage by dissociation of carbon bonds around the cluster. In particular, the translation of the cluster from outside to inside of the cage corresponds to the observed transitions from the patched heterofullerene to the EMF under electron irradiation (Figs. 4e and f) and from the fullerene with the attached cluster to the patched heterofullerene after heat treatment (Fig. 4d). The carbon nanotube wall prevents motion of the nickel cluster out of the fullerene cage. Thus the carbon nanotube provides some assistance to the formation of EMFs versus patched heterofullerenes when compared with the same transformations in vacuum and explains the observed increase in EMF yield inside the carbon nanotube compared to in vacuum.

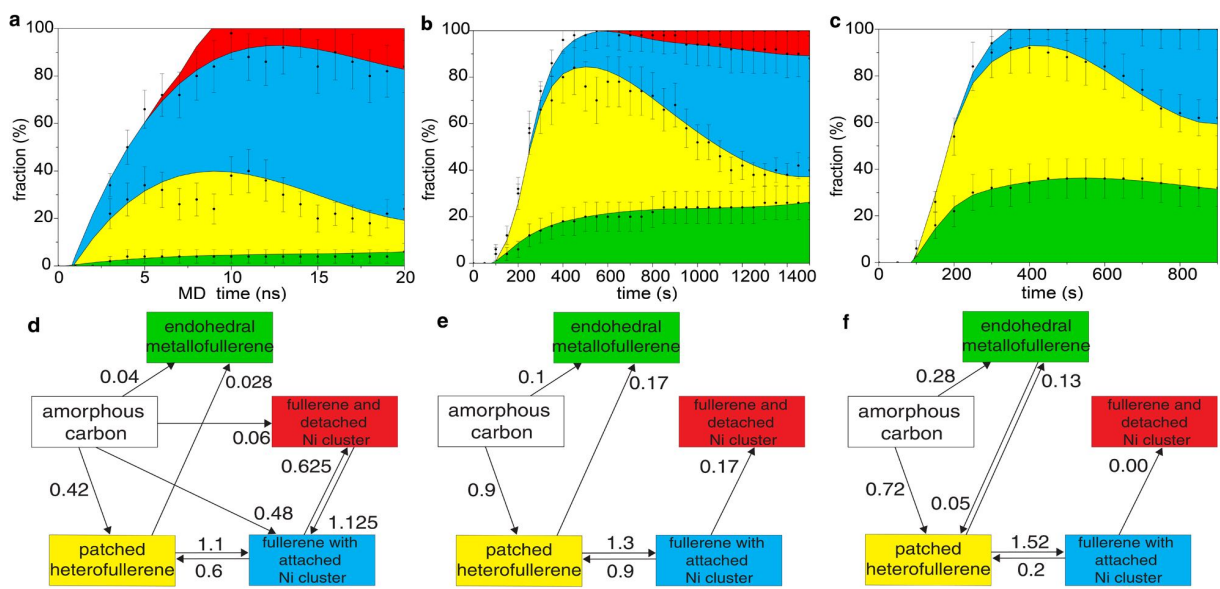

**Figure 4.** Time dependences of fractions of different nanoobjects and number of transitions between them observed in molecular dynamics simulations. (a), (b) and (c), Calculated time dependences of fractions of different nanoobjects: nickel cluster surrounded by amorphous carbon (white), endohedral metallofullerene (green), patched heterofullerene (yellow), fullerene with attached nickel cluster (blue), fullerene and detached nickel cluster (red); (d), (e) and (f), flowcharts showing transitions between the nanoobjects. The conditions considered are (a,d) heat treatment inside the carbon nanotube; (b, e) electron beam irradiation in vacuum; (c, f) electron beam irradiation inside the carbon nanotube. The calculated average number of transitions per single structure of a given type is indicated.

In summary, the synthesis of EMFs containing clusters of up to 50-70 nickel atoms has been demonstrated. EMFs formation is initiated by electron irradiation in AC-HRTEM of the starting nickel cluster surrounded by amorphous carbon or attached to a graphene flake in the interior of a carbon nanotube. Importantly to date it has not been possible to synthesise EMFs with transition metal atoms using traditional EMF synthetic methods, however our proposed method can potentially be extended to other late transition metal atoms. Moreover EMFs obtained up to now at high temperature in the gas phase via the modified arc discharge method have had a maximum of 4-7 atoms of different metal compounds inside the fullerene cage[7,8,9]. Separately, multicage carbon nanoparticles with a metal core of 5-10 nm in diameter have been obtained by high temperature treatment via laser irradiation[39]. Here we illustrate a methodology that fills the gap between these nanoobjects and enables the formation of EMFs with tens of metal atoms inside. It should also be emphasized that the formation mechanism of nickel cluster EMFs achieved at room temperature is principally different from both the assembly of conventional EMFs in the gas phase or formation of multicage carbon nanoparticles with a metal core which are made via a dissolution-precipitation mechanism. In both of the latter methods formation of metal-carbon nanostructures takes place in an open system with an external carbon source. In contrast, in our method the system should be isolated, so that carbon supply from outside of the system is insignificant during the transformation. Such lack of carbon is the main reason of EMF instead of carbon nanotube formation. Another essential feature of the proposed method is that the number of heavy metal atoms is conserved during the total transformation of the initial carbon system into a closed fullerene cage as a result of electron impacts. This is very different to the formation of multicage carbon nanoparticles with the metal core, where merging of nanoparticles was observed[39]. Thus our method enables us to select a specific metal cluster (given number of atoms of different metals), surround it with a carbon nanostructure, and then synthesize a EMF with complete confidence of the composition of the final internal cluster.

We have also explored the formation mechanism of EMFs under electron irradiation, revealing the separate roles of electron irradiation and carbon nanotube templating within the formation of EMFs by considering three specific cases of treatment of nickel clusters surrounded by amorphous carbon; electron irradiation inside a carbon nanotube, under electron irradiation in vacuum and heat treatment inside a carbon nanotube, using molecular dynamics simulations with the CompuTEM algorithm[27,28]. Whilst the formation of EMFs is observed for all three simulated cases, the highest yield of EMFs takes place under electron irradiation inside carbon nanotubes. Thus both electron irradiation and carbon nanotube promote the EMF formation. Since the yield of EMFs under electron irradiation is several times greater than under heat treatment the role of electron irradiation appears crucial. This is related to the fact that under electron irradiation the energy transmitted to the system is mostly transferred to the carbon atoms and, therefore, this energy is almost exclusively spent on the transformation of the carbon nanostructure into the fullerene cage, while during heat treatment some part of the energy provided is consumed on driving the metal through the forming fullerene cage and cluster desorption. The carbon nanotube is demonstrated to act as a very narrow container which prevents motion of the nickel cluster out of the fullerene cage and its subsequent desorption. However the yield of EMFs under electron irradiation inside carbon nanotubes is only slightly less than 50 % larger than that of the same numerical experiment in vacuum. This open up possibilities for quantity production of new types of EMFs and heterofullerenes in vacuum or buffer gas or on a surface by electron irradiation or heat treatment of initial carbon-metal nanostructures if merging of the forming EMFs or heterofullerenes can be prevented. This process is analogous to the graphene flake to fullerene transformation under electron irradiation[10] which have been observed on graphite surface. The methods of synthesis of nanometer-sized graphene flakes are now actively elaborated (see Ref. 40 for a review). Deposition of such flakes and metal clusters can be considered as one of the possible ways to obtain initial carbon-metal nanostructures on a surface which can be transformed into EMFs and heterofullerenes under electron irradiation. Since EMFs can easier diffuse on the surface after their formation in comparison with initial carbon-metal nanostructures and thus can leave the area of synthesis such diffusion can be used for their separation. We believe that our study not only sheds light on the mechanisms of the carbon shell formation around metal atoms, but it can potentially stimulate the research into new and more efficient methods of EMF synthesis.

AUTHOR INFORMATION

**Corresponding Author**

*Address correspondence to liv_ira@hotmail.com, popov-isan@mail.ru

ACKNOWLEDGMENT

A.S.S., I.V.L., A.A.K., and A.M.P. acknowledge the Russian Foundation of Basic Research (14-02-00739-a). The authors also acknowledge computational time on the Supercomputing Center of Lomonosov Moscow State University and the Multipurpose Computing Complex NRC "Kurchatov Institute". I.V.L. acknowledges the financial support from Grupos Consolidados UPV/EHU del Gobierno Vasco (IT578-13) and EU-H2020 project "MOSTOPHOS" (n. 646259). A.N.K., T.W.C. and R.L.M. acknowledge the ERC and EPSRC for financial support, and the Nanoscale & Microscale Research Centre (nmRC), University of Nottingham, for access to instrumentation. T.Z., J.B. and U.K. gratefully acknowledge the funding by the DFG (German Research Foundation) and the Ministry of Science, Research and the Arts (MWK) of Baden Wuerttemberg in the framework of the SALVE (Sub Angstrom Low-Voltage Electron Microscopy) project.

REFERENCES

1. Cong, H.; Yu, B.; Akasaka, T.; Lu, X. Endohedral Metallofullerenes: An Unconventional Core–Shell Coordination Union. *Coordination Chemistry Reviews* **2013,** *257*, 2880–2898.

2. Popov, A. A.; Yang, S.; Dunsch, L. Endohedral Fullerenes. *Chem. Rev.* **2013,** *113*, 5989–6113.

3. Yang, S.; Wang, C.-R. *Endohedral Fullerenes: From Fundamentals to Applications;* World Scientific: Singapore, 2014.

4. Stevenson, S.; Rice, G.; Glass, T.; Harich, K.; Cromer, F.; Jordan, M. R.; Craft, J.; Hadju, E.; Bible, R.; Olmstead, M. M.; Maitra, K.; Fisher, A. J.; Balch, A. L.; Dorn, H. C. Small-Bandgap Endohedral Metallofullerenes in High Yield and Purity. *Nature* **1999,** *401,* 55–57.

5. Wang, C.-R.; Kai, T.; Tomiyama, T.; Yoshida, T.; Kobayashi, Y.; Nishibori, E.; Takata, M.; Sakata, M.; Shinohara, H. A Scandium Carbide Endohedral Metallofullerene: $(Sc_2C_2)@C_{84}$. *Angew. Chem. Int. Ed.* **2001**, *40*, 397–399.

6. Saunders, M.; Jiménez-Vázquez, H. A.; Cross, R. J.; Poreda, R. J. Stable Compounds of Helium and Neon: He@C60 and Ne@C60. *Science* **1993**, *259*, 1428–1430.

7. Komatsu, K.; Murata, M.; Murata, Y. Encapsulation of Molecular Hydrogen in Fullerene $C_{60}$ by Organic Synthesis. *Science* **2005**, *307*, 238–240.

8. Kurotobi, K.; Murata, Y. A Single Molecule of Water Encapsulated in Fullerene $C_{60}$. *Science* **2011**, *333*, 613–616.

9. Mercado, B. Q.; Olmstead, M. M.; Beavers, C. M.; Easterling, M. L.; Stevenson, S. M.; Mackey, M. A.; Coumbe, C. E.; Phillips, J. D.; Phillips, J. P.; Poblet, J. M.; Balch, A. L. A Seven Atom Cluster in a Carbon Cage, the Crystallographically Determined Structure of $Sc_4(\mu_3\text{-}O)_3@I_h\text{-}C_{80}$. *Chem. Commun.* **2010**, *46*, 279–281.

10. Chuvilin, A.; Kaiser, U.; Bichoutskaia, E.; Besley, N. A.; Khlobystov, A. N. Direct Transformation of Graphene to Fullerene. *Nat. Chem.* **2010**, *2*, 450–453.

11. Helveg, S.; Lopez-Cartes, C.; Sehested, J.; Hansen, P. L.; Clausen, B. S.; Rostrup-Nielsen, J. R.; Abild-Pedersen, F.; Norskov, J. K. Atomic-Scale Imaging of Carbon Nanofibre Growth. *Nature* **2004**, *427*, 426–429.

12. Chamberlain, T. W.; Meyer, J. C.; Biskupek, J.; Leschner, J.; Santana, A.; Besley, N. A.; Bichoutskaia, E.; Kaiser, U.; Khlobystov, A. N. Reactions of the Inner Surface of Carbon Nanotubes and Nanoprotrusion Processes Imaged at the Atomic Scale. *Nat. Chem.* **2011**, *3*, 732–737.

13. Zoberbier, T.; Chamberlain, T. W.; Biskupek, J.; Kuganathan, N.; Eyhusen, S.; Bichoutskaia, E.; Kaiser, U.; Khlobystov, A. N. Interactions and Reactions of Transition Metal Clusters with the Interior of Single-Walled Carbon Nanotubes Imaged at the Atomic Scale. *J. Am. Chem. Soc.* **2012**, *134*, 3073–3076.

14. Koshino, M.; Niimi, Y.; Nakamura, E.; Kataura, H.; Okazaki, T.; Suenaga, K.; Iijima, S. Analysis of the Reactivity and Selectivity of Fullerene Dimerization Reactions at the Atomic Level. *Nat. Chem.* **2010**, *2*, 117–124.

15. Lebedeva, I. V.; Chamberlain, T. W.; Popov, A. M.; Knizhnik, A. A.; Zoberbier, T.; Biskupek, J.; Kaiser, U.; Khlobystov, A. N. The Atomistic Mechanism of Carbon Nanotube Cutting Catalyzed by Nickel under an Electron Beam. *Nanoscale* **2014,** *6*, 14877–14890.

16. Yue, Y.; Yuchi, D.; Guan, P.; Xu, J.; Guo, L. Liu, J. Atomic Scale Observation of Oxygen Delivery During Silver–Oxygen Nanoparticle Catalyzed Oxidation of Carbon Nanotubes. *Nat. Commun.* **2016,** *7*, 12251.

17. Biskupek, J.; Hartel, P.; Haider, M.; Kaiser, U. Effects of Residual Aberrations Explored on Single-Walled Carbon Nanotubes. *Ultramicroscopy* **2010,** *116*, 1–7.

18. http://www.qstem.org, QSTEM, 2013.

19. Zoberbier, T.; Chamberlain,T. W.; Biskupek, J.; Suyetin, M.; Majouga, A. G.; Besley, E.; Kaiser, U.; Khlobystov, A. N. Investigation of the Interactions and Bonding between Carbon and Group VIII Metals at the Atomic Scale. *Small* **2016**, 1649–1657.

20. Bandow, S.; Takizawa, M.; Kato, H.; Okazaki, T.; Shinohara, H.; Iijima, S. Smallest Limit of Tube Diameters for Encasing of Particular Fullerenes Determined by Radial Breathing Mode Raman Scattering. *Chem. Phys. Lett.* **2001**, *347*, 23–28.

21. Dunlap, B. I.; Zopeb, R. R. Efficient Quantum-Chemical Geometry Optimization and the Structure of Large Icosahedral Fullerenes. *Chem. Phys. Lett.* **2006**, *422*, 451–454.

22. Lamb, L. D.; Huffman, D. R.; Workman, R. K.; Howells, S.; Chen, T.; Sarid, D.; Ziolo, R. F. Extraction and STM Imaging of Spherical Giant Fullerenes. *Science* **1992**, *255*, 1413–1416.

23. Beavers, C. M.; Jin, H.; Yang, H.; Wang, Z.; Wang, X.; Ge, H.; Liu, Z.; Mercado, B. Q.; Olmstead, M. M.; Balch, A. L. Very Large, Soluble Endohedral Fullerenes in the Series $La_2C_{90}$ to $La_2C_{138}$: Isolation and Crystallographic Characterization of $La_2@D_5(450)$-$C_{100}$. *J. Am. Chem. Soc.* **2011**, *133*, 15338–15341.

24. Yamaguchi, Y.; Maruyama, S. A Molecular Dynamics Study on the Formation of Metallofullerenes. *Eur. Phys. J. D* **1999**, *9*, 385–388.

25. Deng, Q.; Heine, T.; Irle, S.; Popov, A. A. Self-Assembly of Endohedral Metallofullerenes: A Decisive Role of Cooling Gas and Metal-Carbon Bonding. *Nanoscale* **2016**, *8*, 3796–3808.

26. Mulet-Gas, M.; Abella, L.; Dunk, P. W.; Rodríguez-Fortea, A.; Kroto H. W.; Poble J. M. Small Endohedral Metallofullerenes: Exploration of the Structure and Growth Mechanism in the $Ti@C_{2n}$ (2n = 26–50) Family. *Chem. Sci.* **2015**, *6*, 675–686.

27. Santana, A.; Zobelli, A.; Kotakoski, J.; Chuvilin, A.; Bichoutskaia, E. Inclusion of Radiation Damage Dynamics in High-Resolution Transmission Electron Microscopy Image Simulations: the Example of Graphene. *Phys. Rev. B* **2013,** *87*, 094110.

28. Skowron, S. T.; Lebedeva, I. V.; Popov, A. M.; Bichoutskaia, E. Approaches to Modelling Irradiation-Induced Processes in Transmission Electron Microscopy. *Nanoscale* **2013,** *5*, 6677–6692.

29. Sinitsa, A. S.; Lebedeva, I. V.; Knizhnik, A. A.; Popov, A. M.; Skowron, S. T.; Bichoutskaia, E. Formation of Nickel–Carbon Heterofullerenes under Electron Irradiation. *Dalton Trans.* **2014,** *43*, 7499–7513.

30. Cornell, W. D.; Cieplak, P.; Bayly, C. I.; Gould, I. R.; Merz, K. M.; Ferguson, D. M.; Spellmeyer, D. C.; Fox, T.; Caldwell, J. W.; Kollman, P. A. A Second Generation Force field for the Simulation of Proteins, Nucleic Acids, and Organic Molecules. *J. Am. Chem. Soc.* **1995,** *117*, 5179–5197.

31. Mittendorfer, F.; Garhofer, A.; Redinger, J.; Klimeš, J.; Harl, J.; Kresse, G. Graphene on Ni(111): Strong Interaction and Weak Adsorption. *Phys. Rev. B* **2011,** *84*, 201401.

32. Olsen, T.; Thygesen, K. S. Random Phase Approximation Applied to Solids, Molecules, and Graphene-Metal Interfaces: from Van der Waals to Covalent Bonding. *Phys. Rev. B* **2013,** *87*, 075111.

33. Silvestrelli, P. L.; Ambrosetti, A. Van der Waals Corrected DFT Simulation of Adsorption Processes on Transition-Metal Surfaces: Xe and Graphene on Ni(111). *Phys. Rev. B* **2015,** *91*, 195405.

34. http://www.kintechlab.com, Kintech Lab, 1998 – 2017.

35. Berendsen, H. J. C.; Postma, J. P. M.; van Gunsteren, W. F.; DiNola, A.; Haak, J. R. Molecular Dynamics with Coupling to an External Bath. *J. Chem. Phys.* **1984,** *81*, 3684–3690.

36. Irle, S., Zheng, G., Wang, Z. & Morokuma, K. The $C_{60}$ formation puzzle "solved": QM/MD simulations reveal the shrinking hot giant road of the dynamic fullerene self-assembly mechanism. *J. Phys. Chem. B* **2006** *110*, 14531–14545.

37. Chuvilin, A.; Bichoutskaia, E.; Gimenez-Lopez, M. C.; Chamberlain, T. W.; Rance, G. A.; Kuganathan, N.; Biskupek, J.; Kaiser, U.; Khlobystov, A. N. Self-Assembly of a Sulphur-Terminated Graphene Nanoribbon within a Single-Walled Carbon Nanotube, *Nat. Mat.* **2011,** *10*, 687–692.

38. Lebedeva, I. V.; Knizhnik, A. A.; Popov, A. M.; Potapkin, B. V. Ni-Assisted Transformation of Graphene Flakes to Fullerenes. *J. Phys. Chem. C* **2012,** *116*, 6572–6584.

39. Chandrakumar, K. R. S.; Readle, J. D.; Rouleau, C.; Puretzky, A.; Geohegan, D. B.; More, K.; Krishnan, V.; Tian, M.; Duscher, G.; Sumpter, B.; Irle, S.; Morokuma, K. High-Temperature Transformation of Fe-Decorated Single-Wall Carbon Nanohorns to Nanooysters: A Combined Experimental and Theoretical Study. *Nanoscale* **2013,** *5*, 1849–1857.

40. Bacon, M.; Bradley S. J.; Nann, T. Graphene Quantum Dots. *Part. Part. Syst. Charact.* **2014**, *31*, 415–428.

# Supporting information

**Formation of nickel clusters wrapped in carbon cages: towards new endohedral metallofullerene synthesis**

*Alexander S. Sinitsa[1], Thomas W. Chamberlain[2], Thilo Zoberbier[3], Irina V. Lebedeva[4], Andrey M. Popov[5,*], Andrey A. Knizhnik[1,6], Robert L. McSweeney[7], Johannes Biskupek[3], Ute Kaiser[3], Andrei N. Khlobystov[7]*

[1] National Research Centre "Kurchatov Institute", Kurchatov Square 1, Moscow 123182, Russia
[2] Institute of Process Research and Development, School of Chemistry, University of Leeds, Leeds, LS2 9JT, UK
[3] Group of Electron Microscopy of Materials Science, Central Facility for Electron Microscopy, Ulm University, Albert Einstein Allee 11, Ulm 89081, Germany
[4] Nano-Bio Spectroscopy Group and ETSF, Universidad del País Vasco, CFM CSIC-UPV/EHU, San Sebastian 20018, Spain
[5] Institute for Spectroscopy of Russian Academy of Sciences, Fizicheskaya Street 5, Troitsk, Moscow 108840, Russia
[6] Kintech Lab Ltd., 3rd Khoroshevskaya Street 12, Moscow 123298, Russia
[7] School of Chemistry, University of Nottingham, University Park, Nottingham NG7 2RD, UK

## AC-HRTEM time series

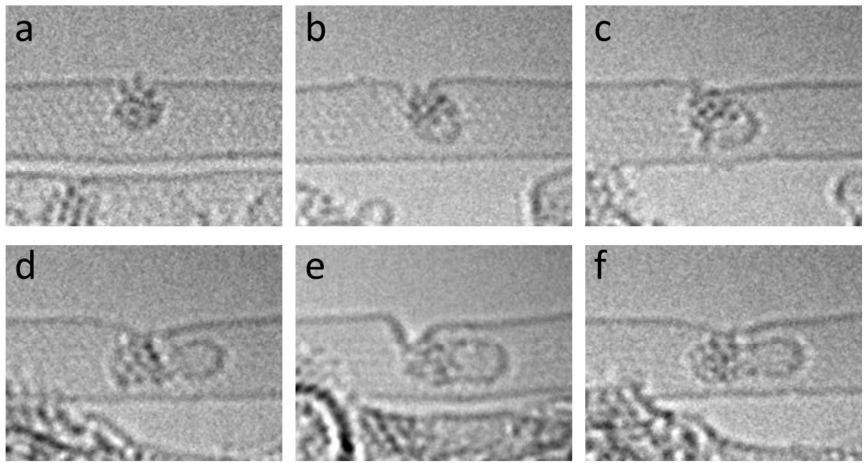

**Figure S1.** Structure evolution under 80 keV electron beam irradiation: time series of AC-HRTEM images (a-f) showing structural transformations initiated by the electron beam. Experimental detail: total time 770 s, cumulative dose $3.4 \cdot 10^9$ e$^-$/nm$^2$. Scale bar is 1 nm.

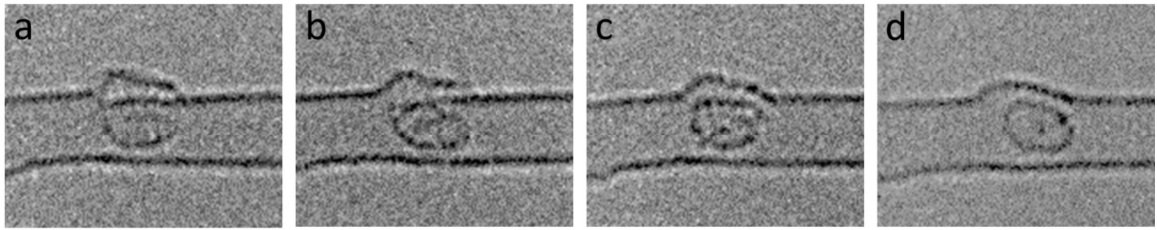

**Figure S2.** Structure evolution under 80 keV electron beam irradiation: time series of AC-HRTEM images (a-d) showing structural transformations initiated by the electron beam. Experimental detail: total time 63 s, cumulative dose $1.2 \cdot 10^8$ e$^-$/nm$^2$. Scale bar is 1 nm.

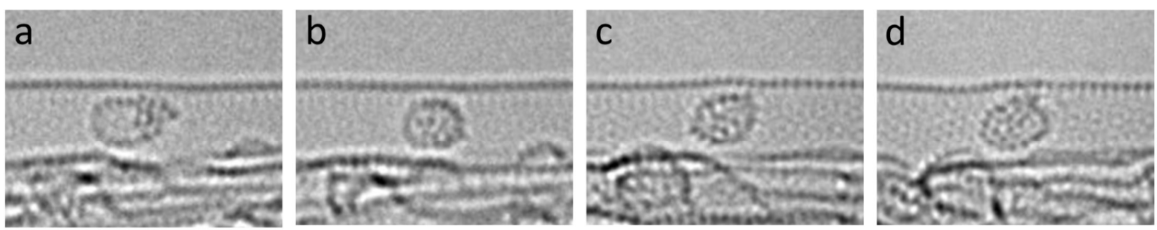

**Figure S3.** Structure evolution under 80 keV electron beam irradiation: time series of AC-HRTEM images (a-d) showing structural transformations initiated by the electron beam. Experimental detail: total time 480 s, cumulative dose $5.0 \cdot 10^9$ e$^-$/nm$^2$. Scale bar is 1 nm.

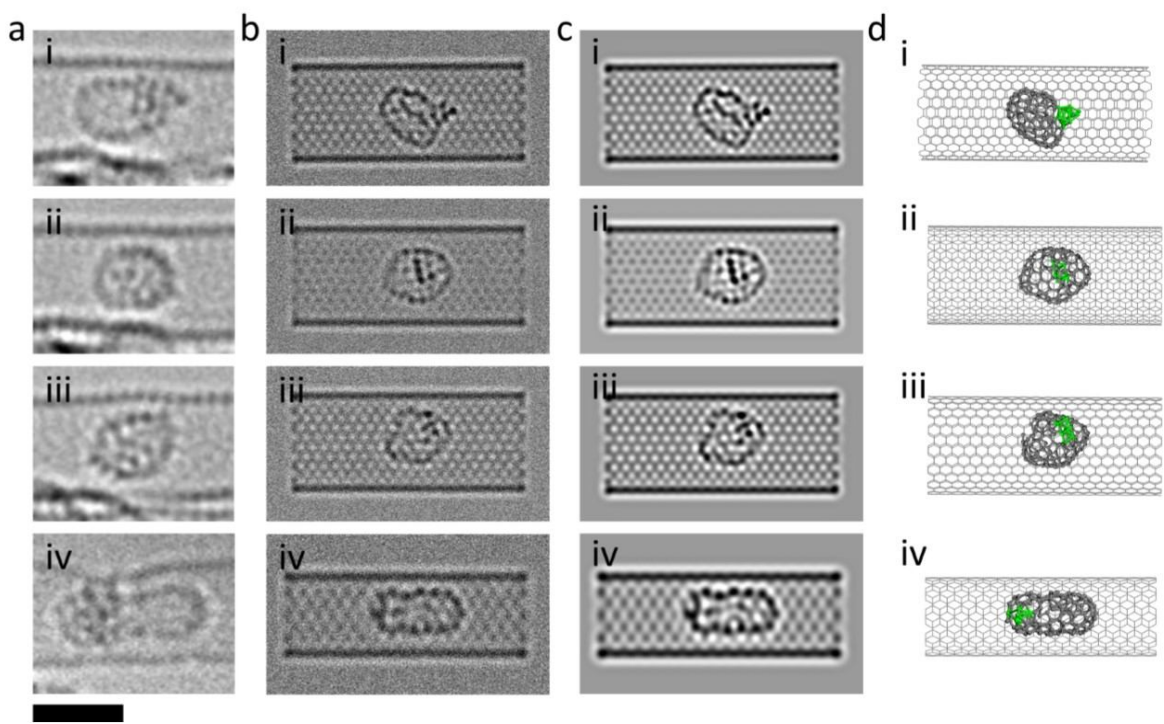

**Figure S4.** Theoretical simulation of the structures observed in the NiNPs@SWNT sample by AC-HRTEM at 80 keV: (a)(i-iv) AC-HRTEM images of nickel clusters interacting with carbon fullerene structures show strong correlation with the contrast observed in simulated TEM images generated using limited (b)(i-iv) and infinite (c)(i-iv) doses of 80 keV electrons applied to the corresponding structural models (d)(i-iv). Scale bar is 1 nm.

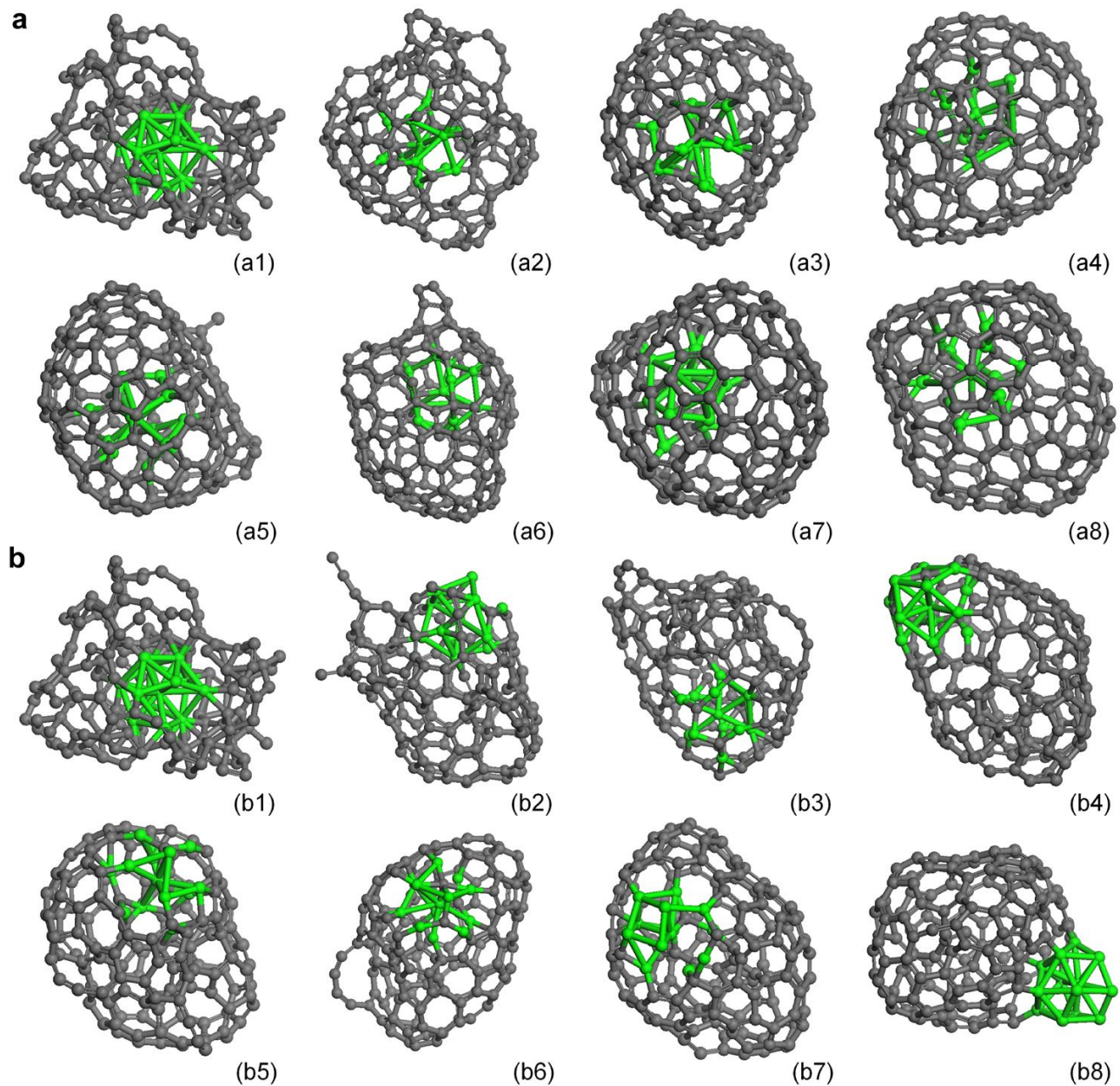

**Figure S5.** Structure evolution in molecular dynamics simulations in vacuum: simulated structure evolution of the amorphous carbon with the nickel cluster attached in vacuum under electron irradiation in HRTEM: (a) transformation into an endohedral metallofullerene, (a1) 0 s, (a2) 103 s, (a3) 251 s, (a4) 300 s, (a5) 554 s, (a6) 700 s, (a7) 851 s and (a8) 1350 s; (b) into a patched fullerene, (b1) 0 s, (b2) 103 s, (b3) 200 s, (b4) 503 s, (b5) 650 s, (b6) 1000 s, (b7) 1200 s and (b8) 1250 s.

**Note S1: Differentiation of structures in molecular dynamics simulations**

To obtain information on the topology of the composite structures observed in molecular dynamics simulations of the amorphous carbon and an attached nickel cluster, their bond network was analyzed on the basis of the "shortest-path" algorithm[1]. Two carbon atoms are considered as bonded if the distance between them does not exceed 1.8 Å, while for bonded carbon and nickel atoms, this maximal bond length is 2.2 Å. We then divided all of the carbon atoms in 22 types, depending on the number of carbon and nickel neighbours, their local environment and the distribution of non-hexagonal rings.

Our analysis shows that there are two main types of carbon atoms that characterize the shape of the obtained hollow fullerene-like structures, these are the number of two-coordinated carbon atoms in chains, $N_c$, and the number of three-coordinated atoms, $N_{3a}$, belonging to the amorphous domain of the structure. In both of these cases only atoms that are not bound to the nickel cluster are taken into account. Each three-coordinated atom is assigned with a unit vector that is normal to the plane passing through each of the three neighbors of the considered atom. If the angle $\alpha$ between these vectors at two bonded three-coordinated atoms is such that $|cos\alpha| < 0.7$, i.e. the atoms do not lie on a smooth surface, both of these atoms are labelled as belonging to the amorphous domain. Two-coordinated atoms are assumed to be in chains if at least one of the neighbouring carbon atoms is also two- or one-coordinated.

During molecular dynamics runs the number of $N_c$ and $N_{3a}$ atoms decreases gradually or sometimes rather abruptly as the amorphous carbon transforms into a hollow fullerene cage. We suppose that the amorphous carbon structure is transformed irreversibly into a fullerene-like structure when either the number of carbon atoms in chains, $N_c$, or the number of three-coordinated carbon atoms in the amorphous domain, $N_{3a}$, reaches half of their initial values (depending on which of the quantities halves first). Such a decrease demonstrates that the initial irregular amorphous carbon structure ($N_c + N_{3a} \approx N_{tot}^C / 2$, where $N_{tot}^C$ is the total number of carbon atoms) is organized into a more regular fullerene-like structure with a cavity in the center.

To differentiate the fullerene-like structures according to the relative position of the nickel cluster we also analyzed the total number of carbon atoms connected with the nickel cluster, $N_{tot}^{CNi}$, and the average distance between the centers of mass of the carbon atoms initially belonging to the amorphous structure and nickel atoms in the cluster, $r_{CNi} = \left| \mathbf{r}_{cm}^C - \mathbf{r}_{cm}^{Ni} \right|$, where vectors $\mathbf{r}_{cm}^C$ and $\mathbf{r}_{cm}^{Ni}$ describe positions of the centers of mass of the carbon and nickel atoms, respectively. These two quantities allow us to separate the following types of nanoobjects: (1) endohedral metallofullerenes, (2) heterofullerenes with a nickel patch, (3) fullerenes with the nickel cluster attached to the outside and (4) fullerenes and detached nickel clusters. For endohedral metallofullerenes $r_{CNi} < 3$ Å. For fullerenes and detached nickel clusters $r_{CNi} > 8$ Å ($r_{CNi}$ usually rises abruptly at the moment when the nickel cluster detaches and flies away from the fullerene). For heterofullerenes with the nickel patch and fullerenes with the attached nickel cluster, $r_{CNi}$ is between 3 and 8 Å. However, there are more carbon atoms connected with the nickel cluster for heterofullerene structures. In the case when the nickel cluster is attached, the total number of carbon atoms connected with the nickel cluster, $N_{tot}^{CNi}$, is less than 10 (including carbon atoms dissolved in the cluster), while in heterofullerenes it exceeds 10.

Based on these two criteria, the first one to determine the transition from the initial amorphous carbon structure into a fullerene and the second one to differentiate between fullerene-like nanoobjects according to the relative position of the nickel cluster, we follow the transformations of the structure during MD runs. To obtain time dependences of fractions of different structures and the number of transitions between them (Fig. 3) the parameters used in these criteria ($N_c$, $N_{3a}$, $N_{tot}^{CNi}$ and $r_{CNi}$) are averaged over 50 seconds in the simulations under electron irradiation and 1 ns in the simulations under heat treatment.

## References

1. Franzblau, D. C. Computation of Ring Statistics for Network Models of Solids. *Phys. Rev. B* **1991**, *44*, 4925–4930.